\documentclass[a4paper,pra,twocolumn,showpacs,superscriptaddress,groupedaddress]{revtex4}
\usepackage{ae}
\usepackage{physics}
\usepackage[T1]{fontenc}
\usepackage[ansinew]{inputenc}
\usepackage{amsmath}
\usepackage{amssymb}
\usepackage[caption=false]{subfig}
\usepackage{multirow}
\usepackage{array}
\usepackage[]{graphicx}
\usepackage{wrapfig}
\usepackage{makecell}
\usepackage{romanbar}

\usepackage{color}
\usepackage[colorlinks]{hyperref}
\usepackage{lscape}
\begin{document}
\title{Necessary and Sufficient Condition for the Equivalence of Two Pure Multipartite States under Stochastic Local Incoherent Operations and Classical Communications}
\author{Dipayan Chakraborty}
\email{dipayan.tamluk@gmail.com}
\affiliation{Department Of Applied Mathematics, University Of Calcutta, 92, A.P.C. Road, Kolkata 700009,India}
\author{Prabir Kumar Dey}
\email{prabirkumardey1794@gmail.com}
\affiliation{Department Of Applied Mathematics, University Of Calcutta, 92, A.P.C. Road, Kolkata 700009,India}
\author{Nabendu Das}
\email{nabendudas92@gmail.com}
\affiliation{Department Of Applied Mathematics, University Of Calcutta, 92, A.P.C. Road, Kolkata 700009,India}
\author{Indrani Chattopadhyay}
\email{icappmath@caluniv.ac.in}
\affiliation{Department Of Applied Mathematics, University Of Calcutta, 92, A.P.C. Road, Kolkata 700009,India}
\author{Amit Bhar}
\email{bharamit79@gmail.com}
\affiliation{Department of Mathematics, Jogesh Chandra Chaudhuri College, 30, Prince Anwar Shah Road, Kolkata 700033, India}
\author{Debasis Sarkar}
\email{dsarkar1x@gmail.com, dsappmath@caluniv.ac.in}
\affiliation{Department Of Applied Mathematics, University Of Calcutta, 92, A.P.C. Road, Kolkata 700009,India}
\begin{abstract}
Resource theory of quantum coherence originated like entanglement in quantum information theory. However, still now proper classification of quantum states is missing under coherence. In this work, we have provided a classification of states under local incoherent operations. We have succeeded in deriving the necessary and sufficient condition for which two pure multipartite states are equivalent under stochastic local incoherent operations and classical communications (SLICC) and local incoherent operations and classical communications (LICC). In particular, we have succeeded in characterizing three qubit pure states under SLICC. Our result reveals the existence of infinite number of SLICC inequivalent classes for three qubit systems. 
\end{abstract}
\date{\today}
\maketitle

\section{Introduction}
Resource theories offer a powerful framework in quantum information theory for understanding how certain physical properties naturally change within a physical system \cite{1,2,3}. It provides a tool to achieve tasks that are not possible by the laws of classical physics. Quantum resource theory generally consists of two basic elements: (i) free  states and (ii) free operations. Entanglement is one of the most useful resource in quantum information theory where free states are separable states and free operations are local operations alongwith classical communications (in short, LOCC). Several uses of entanglement including teleportation, dense coding \cite{4,5}, quantum computation \cite{5,6}, quantum cryptography\cite{2}, provide us entanglement as a good resource in quantum information processing tasks. \\

Quantum coherence is another important resource which originates from superposition of quantum states \cite{7,8,9,10,11,12,13,14,15,16,17,18,26}. It plays an important role in the study of quantum thermodynamics \cite{19,20,21}, biological systems \cite{22}, etc. Baumgratz et al., proposed the basic notions of incoherent states, incoherent operations and also proposed the necessary conditions that any measure of coherence should satisfy \cite{7}. There are lot of research works done on resource theory of coherence on single qubit systems. For better understanding of resource theoretic aspects of coherence, it is also necessary to study coherence in multipartite scenario. On the basis of classification of multipartite states, one could understand the coherence in multipartite system more profoundly and that could help us to carry out coherence as a resource in implementing  various protocols in multipartite systems. Classification of bipartite systems in coherence resource theory was proposed recently \cite{27}. In our work, we shall discuss on classification of multipartite system in general and in particular, pure three qubit states under local incoherent operations. \\

In resource theory of entanglement a complete classification of pure three qubit states with respect to stochastic local operation with classical communication(SLOCC) were proposed \cite{23}. In this classification, conversion of states were done through SLOCC, i.e., through LOCC but without restricting that it has to be done with certainty. This classification contains separable states, biseparable states and two genuine three qubit entangled states namely GHZ and W states. Conversion of states between two  different classes is not possible via SLOCC. In our work, we will classify pure three qubit states with respect to stochastic local incoherent operation with classical communication(SLICC). We will first derive necessary and sufficient condition for equivalence of two states under SLICC. Thereafter, we will use this result to classify pure three qubit states with respect to SLICC. We will provide an interesting result that there exists infinite number of SLICC inequivalent class under SLICC in contrast with only six SLOCC inequivalent classes exists for pure three qubit states. 

\section{Review on resource theory of Coherence}
Characterization of resource theory of coherence is based on mainly two essential ingredients, incoherent states as free states and incoherent operations as free operations. \\

\textbf{ Incoherent states:} To define incoherent states, we first fix a specific basis \{$\ket{i},i=1,....,d$\} of the Hilbert space representing a quantum system. The set of incoherent states $ I $ are the all density matrices which are diagonal in this basis \cite{7}: $$\rho=\sum_i p_i\ket{i}\bra{i}$$ 

Definition of incoherent operations in resource theory of coherence are not unique. Two important incoherent operations are,

\textbf{Incoherent operations:} Incoherent operations \cite{7} are characterized as the set of trace preserving completely positive maps  admitting a set of Kraus operators $\lbrace K_n \rbrace$ such that $\sum_n K_n^\dagger K_n=I$ and, for all n and $\rho\in I$, $$\frac{K_n\rho K_n^\dagger}{Tr[K_n\rho K_n^\dagger]}\in I$$ 
From the definition of incoherent operations, it is clear that for any possible outcome, coherence can never be generated from incoherent states via this operation, not even probabilistically.

Kraus  operators representing incoherent operations has the form $K_n=\sum_n c_n\ket{f(n)}\bra{n}$, where $\ket{f(n)}$ is many to one function from basis set onto itself.\\
 
\textbf{Strictly incoherent operations:} Strictly incoherent operations \cite{17,18} are represented by set of completely positive, trace preserving maps having Kraus operator representation $\lbrace K_n \rbrace_n$ such that $$ K_n\bigtriangleup(\rho)K_n^\dagger = \bigtriangleup(K_n \rho K_n^\dagger) \; \;  \forall  n  , \forall \; \rho ,$$ where $\Delta$ denotes completely dephasing map.

Kraus  operator representing strictly incoherent operations has the form $K_n=\sum_n c_n
\ket{\pi(n)}\bra{n}$, where $\ket{\pi(n)}$ is  permutation from basis set onto itself. The quantification procedure is based on the following notions.\\

\textbf{Measure of coherence:} Any functional $C$ which maps the set of states to the set of non negative real numbers should satisfy following properties in order to be a proper coherence measure \cite{7}:\\
 (C1)Non-negativity: $$C(\rho)\geq 0$$ in general, with equality if and only if $\rho$ is incoherent.\\
 (C2) Monotonicity: $C$ does not increase under the action of incoherent operations, i.e., $$C(\Lambda[\rho])\leq C(\rho)$$ for any incoherent operation $\Lambda$.\\
 (C3) Strong monotonicity: $C$ does not increase on average under selective incoherent operations, i.e., $$\sum_iq_iC(\sigma_i)\leq C(\rho)$$ with probabilities $q_i=Tr[K_i\rho K_i^\dagger]$, post measurement states $\sigma_i=\frac{K_i\rho K_i^\dagger}{q_i}$, and incoherent Kraus operators $K_i$.\\
 (C4) Convexity: $C$ is a convex function of the state, i.e., 
 $$ \sum_i p_iC(\rho_i)\geq C(\sum_i p_i\rho_i)$$

\textbf{Incoherent operations in multipartite case:} Now, it comes to identify the free operations in multiparty systems. The framework of local operation and classical communications (LOCC) is an important part of resource theory of entanglement as it is used in various quantum information processing tasks like, teleportation, state transformations, etc. In the LOCC protocol, multiple parties who are spatially separated from each other are allowed to perform only local operations on their subsystems and they are allowed to communicate with each other via classical communication channels. It is very difficult to represent this protocol mathematically as LOCC operation can include arbitrary number of classical communications.
Now, if in the above protocol, all the parties are allowed to perform only incoherent operations then corresponding protocol is known as local incoherent operation with classical communications(LICC) \cite{11,12}.  Like LOCC it is also very difficult to represent whole LICC. operation mathematically.
 
Stochastic LICC (SLICC) operations describe state which can be interconverted by LICC non deterministically but with a non zero probability of success.\\

\section{Necessary and sufficient condition for equivalence of two multipartite pure states under SLICC and LICC}
 
To focus our main results we adopt some important consequences from multipartite entanglement.\\
 
\textbf{Lemma 1:}\cite{23} If the vectors $\ket{\psi}$, $\ket{\phi}$ $\in$ $\mathcal{H}_A \otimes \mathcal {H}_B \otimes .........\otimes \mathcal{H}_N$ are connected by a local operator as $\ket{\phi}=A\otimes B \otimes .........\otimes N \ket{\psi} $, then the local ranks satisfy $r(\rho_k^{\psi})\geqslant r(\rho_k^{\phi})$, k=A,B,......,N.\\
 
\textbf{Lemma 2:}\cite{23} Two pure states of a multipartite system are equivalent under SLOCC if and only if they are related by a local, invertible operator.\\
 
Above lemma gives necessary and sufficient condition for equivalence of two multipartite states under SLOCC. Now, we shall use this lemma to prove necessary and sufficient condition for equivalence of two multipartite states under SLICC. The proof is very similar to above lemma.\\
 
\textbf{Lemma 3:} Two pure states of a multipartite system are equivalent under SLICC if and only if they are related by local, strictly incoherent operators $A,B,...,N$ such that
$$\ket{\phi}=A\otimes B\otimes ...\otimes N \ket{\psi}$$
$\textit{Proof:}$ Let the given condition holds, i.e., $\ket{\phi}=A\otimes B\otimes ...\otimes N \ket{\psi},$  where A,B,...,N are local incoherent, invertible operators, then we can transform $\ket{\psi}$ into $\ket{\phi}$ using local protocol with finite probability of success. For this protocol  Alice need to apply POVM defined by operators, $\sqrt{p_A}$A and $\sqrt{I_A-p_A A^\dagger A}$ where $p_A \leqslant 1$ is a possible weight such that $p_A A^\dagger A \leqslant I_A$ and similarly for the rest of the parties.\\
Then this protocol converts $\ket{\psi}$ to $\ket{\phi}$ successfully with probability $p_Ap_B...p_N$. If in addition, A,B,...,N are invertible operators then obviously 
$$\ket{\psi}=A^{-1}\otimes B^{-1}\otimes ...\otimes N^{-1} \ket{\phi}.$$
Therefore, the conversation can be reversed locally.
For the converse part, let $\ket{\psi}$ and $\ket{\phi}$ are equivalent under SLICC. Then a local, incoherent  operator relates them. We need to proof this operator can always be chosen to be invertible. \\
For simplicity, we first assume that  $\ket{\psi}$ and $\ket{\phi}$ are related by a local, incoherent operator acting non-trivially only in Alice's part, i.e.,
$$\ket{\phi}=A\otimes I_{B...N}\ket{\psi}.$$

We can consider Schmidt decomposition of states with respect to part A and B,...,N.
$$\ket{\psi}=\sum_{i=1}^{n_{\psi}} \sqrt{\lambda _i^{\psi}}\ket{i}\ket{\tau_i}$$
$$\ket{\phi}=\sum_{i=1}^{n_{\phi}} \sqrt{\lambda _i^{\phi}}(U_A\ket{i})\ket{\tau_i}$$   
where $\lambda _i^{\psi}, \lambda _i^{\phi}> 0$ for all i.\\
Since, $\ket{\psi}$ and $\ket{\phi}$ are equivalent under SLICC, from previous lemma, we have $n_{\psi}=n_{\phi}$. Since each party are restricted to perform incoherent operations only, we can relate two local Schmidt bases in Alice's part by an incoherent, local unitary $U_A$. Here, $\{\ket{i}\}_{i=1}^n \in \mathcal{H}_A$ and $\{\ket{\tau_i}\}_{i=1}^n \in \mathcal{H}_B\otimes ...\otimes \mathcal{H}_N$.\\
Thus the operator A in the first equation must be $A=U_A (A_1 +A_2)$ \\
where $U_A$ is an incoherent unitary and 
$$A_1 =\sum_{i=1}^{n_{\psi}} \sqrt{\frac{\lambda_i ^{\phi}}{\lambda_i ^{\psi}}} \ket{i}\bra{i}$$
$$A_2 =\sum_{i=n_{\psi}+1}^{n} \ket{\mu_i }\bra{i}.$$
Clearly, $A_1$ is incoherent and $\{\ket{\mu_i}\}$ are arbitrary unnormalized vectors. $\{\ket{\mu_i}\}$ play no role here as
 $A_2\otimes I_{B...N}\ket{\psi}=0$. So to make $A_2$ also incoherent, we can write $A_2$ as 
$$A_2 =\sum_{i=n_{\psi}+1}^{n} \ket{i}\bra{i}.$$ 
So, $(A_1 +A_2)$ is a diagonal matrix with full rank, which implies it is invertible and an incoherent operator.
Therefore, $A=U_A (A_1 +A_2)$ becomes an invertible incoherent operator.\\
The general case would correspond to composing operators $A\otimes I_{B...N}$ with operator $I_A \otimes B\otimes I_{C...N}$ and similarly for the rest of the parties. The above argument should then be applied sequentially to each party individually.\\
Hence, it is proved that if $\ket{\psi}$ and $\ket{\phi}$ are equivalent under SLICC, then they are related by local, incoherent, invertible operators.\\
The above condition is equivalent to the states $\ket{\psi}$ and $\ket{\phi}$ are equivalent under SLICC, if and only if they are related by local, strictly incoherent operator(SIO).\\

\textbf{Lemma 4:} Two pure states $\ket{\psi}$ and $\ket{\phi}$ of a multipartite system are equivalent under LICC if and only if they are related by local, incoherent, invertible and unitary operators $A,B,...,N$ such that
$$\ket{\phi}=A\otimes B\otimes ...\otimes N \ket{\psi}.$$

Proof: Using LICC protocol we can convert a state $\ket{\psi}$ into a state $\ket{\phi}$ with certainty, i.e., this is similar like SLICC protocol but state conversion are done with unit probability. If we want to convert a state $\ket{\psi}$ into $\ket{\phi}$ with unit probability then we have to choose operation such that trace is preserved, i.e., operation must satisfy completeness condition. In the above lemma, if we choose three operators A,B and C such that they satisfy completeness condition, i.e., $A^{\dagger}A\otimes B^{\dagger}B\otimes C^{\dagger}C = I$. This is possible if and only if we choose A,B,C unitary operators, i.e., $A^{\dagger}A=I$, $B^{\dagger}B=I$, $C^{\dagger}C=I$. This is clearly a trace preserving operation. Thus, we can always transfer two pure states with unit probability under LICC.

Now our aim is to classify pure three qubit states with respect to SLICC using above lemma.
For example, equivalence of two pure three qubit states $\ket{\psi}$ and $\ket{\phi}$ under SLICC could be done if we choose three local, strictly incoherent operators $ A,B,C $ as follows:
 \[ 1)\; \; \left( \begin{array}{cc}
a_1 & 0 \\
0 & a_2
\end{array}  \right)  ,
\left( \begin{array}{cc}
b_1 & 0 \\
0 & b_2
\end{array} \right)  , 
\left( \begin{array}{cc}
c_1 & 0 \\
0 & c_2
\end{array} \right)  
\]
\[ 2)\; \; \left( \begin{array}{cc}
a_1 & 0 \\
0 & a_2
\end{array}  \right)  ,
\left( \begin{array}{cc}
b_1 & 0 \\
0 & b_2
\end{array} \right)  , 
\left( \begin{array}{cc}
0 & c_1 \\
c_2 & 0
\end{array} \right)  
\]
\[ 3)\; \; \left( \begin{array}{cc}
a_1 & 0 \\
0 & a_2
\end{array}  \right)  ,
\left( \begin{array}{cc}
0 & b_1 \\
b_2 & 0
\end{array} \right)  , 
\left( \begin{array}{cc}
c_1 & 0 \\
0 & c_2
\end{array} \right)  
\]
\[ 4)\; \; \left( \begin{array}{cc}
0 & a_1 \\
a_2 & 0
\end{array}  \right)  ,
\left( \begin{array}{cc}
b_1 & 0 \\
0 & b_2
\end{array} \right)  , 
\left( \begin{array}{cc}
c_1 & 0 \\
0 & c_2
\end{array} \right)  
\]
\[ 5)\; \; \left( \begin{array}{cc}
0 & a_1 \\
a_2 & 0
\end{array}  \right)  ,
\left( \begin{array}{cc}
0 & b_1 \\
b_2 & 0
\end{array} \right)  , 
\left( \begin{array}{cc}
c_1 & 0 \\
0 & c_2
\end{array} \right)  
\]
\[ 6)\; \; \left( \begin{array}{cc}
0 & a_1 \\
a_2 & 0
\end{array}  \right)  ,
\left( \begin{array}{cc}
b_1 & 0 \\
0 & b_2
\end{array} \right)  , 
\left( \begin{array}{cc}
0 & c_1 \\
c_2 & 0
\end{array} \right)  
\]
\[ 7)\; \; \left( \begin{array}{cc}
a_1 & 0 \\
0 & a_2
\end{array}  \right)  ,
\left( \begin{array}{cc}
0 & b_1 \\
b_2 & 0
\end{array} \right)  , 
\left( \begin{array}{cc}
0 & c_1 \\
c_2 & 0
\end{array} \right)  
\]
\[ 8)\; \; \left( \begin{array}{cc}
0 & a_1 \\
a_2 & 0
\end{array}  \right)  ,
\left( \begin{array}{cc}
0 & b_1 \\
b_2 & 0
\end{array} \right)  , 
\left( \begin{array}{cc}
0 & c_1 \\
c_2 & 0
\end{array} \right)  
\]

If we can relate two pure three qubit states $\ket{\psi}$ and $\ket{\phi}$ by $\ket{\phi}=A\otimes B \otimes C \ket{\psi} $ with $A,B,C$ in any of the above combinations [(1)-(8)] then we can say that two states will lie in same SLICC class.\\

\section{CLASSIFICATION OF PURE THREE QUBIT STATES}

Now, we have provided the complete classification of pure three qubit states with respect to SLICC in tabular format is given below. Here all SLICC inequivalent classes of pure three qubit states are vividly observed. Each state of the table together with its SLICC equivalent states form one SLICC inequivalent class. It is further investigated whether infinite number of SLICC inequivalent classes exist depending on coefficients of the states\\

In the tabular format we shall use the following notations: 
 $\Delta_1=\frac{ad}{bc}$, $\Delta_2=\frac{af}{be}$, $\Delta_3=\frac{af}{ce}$, $\Delta_4=\frac{af}{de}$, $\Delta_5=\frac{ae}{bd}$, $\Delta_6=\frac{af}{dc}$, $\Delta_7=\frac{bf}{cd}$, $\Delta_8=\frac{ag}{ce}$, $\Delta_9=\frac{ae}{cd}$, $\Delta_{10}=\frac{be}{cd}$, $\Delta_{11}=\frac{a^2e}{bcd}$.\\$\Delta_1'=\frac{a'd'}{b'c'}$ and this case will be similar for $\Delta_2'$ to $\Delta_{11}'$.\\
 
$\bullet$ Number of product terms in the state is 1:
\begin{center}
\begin{tabular}{|c|c|c|c|}
\hline
\thead{No.} & \thead{SLICC inequivalet\\states under\\  same no \\of product terms\\ in the state}  & \thead{Existence of \\ infinite SLICC \\inequivalent\\ classes and \\ conditions} \\
\hline
1 & $\ket{000}$ & No\\
\hline

\end{tabular}
\end{center}

$\bullet$ Number of product terms in the state is 2:
\begin{center}
\begin{tabular}{|c|c|c|c|}
\hline
\thead{No.} & \thead{SLICC inequivalet\\states under\\  same no \\of product terms\\ in the state}  & \thead{Existence of \\ infinite SLICC \\inequivalent\\ classes and \\ conditions} \\
\hline
2a & $a\ket{000}+b\ket{001}$ & No\\
\hline
2b & $a\ket{000}+b\ket{010}$ & No\\
\hline
2c & $a\ket{000}+b\ket{011}$ & No\\
\hline
2d & $a\ket{000}+b\ket{100}$ & No\\
\hline
2e & $a\ket{000}+b\ket{101}$ & No\\
\hline
2f & $a\ket{000}+b\ket{110}$ & No\\
\hline
2g & $a\ket{000}+b\ket{111}$ & No\\
\hline
\end{tabular}
\end{center}
%\newpage

$\bullet$ Number of product terms in the state is 3:
\begin{center}
\begin{tabular}{|c|c|c|c|}
\hline
\thead{No.} & \thead{SLICC inequivalet\\states under\\  same no \\of product terms\\ in the state}  & \thead{Existence of \\ infinite SLICC \\inequivalent\\ classes and \\ conditions} \\
\hline
3a & $a\ket{000}+b\ket{001}+c\ket{010}$  & No\\
\hline
3b & $a\ket{000}+b\ket{001}+c\ket{100}$  & No\\

\hline
3c & $a\ket{000}+b\ket{001}+c\ket{110}$  & No\\

\hline
3d & $a\ket{000}+b\ket{011}+c\ket{100}$  & No\\
\hline

3e & $a\ket{000}+b\ket{011}+c\ket{101}$  & No\\

\hline
3f & $a\ket{000}+b\ket{010}+c\ket{100}$  & No\\

\hline
3g & $a\ket{000}+b\ket{010}+c\ket{101}$  & No\\
\hline
\end{tabular}
\end{center}

$\bullet$ Number of product terms in the state is 4:
\begin{center}
\begin{tabular}{|c|c|c|c|}
\hline
\thead{No.} & \thead{SLICC inequivalet\\states under\\  same no \\of product terms\\ in the state}  & \thead{Existence of \\ infinite SLICC \\inequivalent\\ classes and \\ conditions} \\
\\
\hline

4a & \makecell{$a\ket{000}+b\ket{001}+c\ket{010}$\\$+d\ket{100}$} & No\\
\hline

4b & \makecell{$a\ket{000}+b\ket{001}+c\ket{010}$\\$+d\ket{101}$} & No\\
\hline

4c & \makecell{$a\ket{000}+b\ket{001}+c\ket{010}$\\$+d\ket{110}$} & No\\
\hline

4d & \makecell{$a\ket{000}+b\ket{001}+c\ket{010}$\\$+d\ket{111}$} & No\\
\hline

4e & \makecell{$a\ket{000}+b\ket{001}+c\ket{100}$\\$+d\ket{101}$} & No\\
\hline

4f & \makecell{$a\ket{000}+b\ket{001}+c\ket{100}$\\$+d\ket{110}$} & No \\
\hline

4g & \makecell{$a\ket{000}+b\ket{001}+c\ket{100}$\\$+d\ket{111}$} & No\\
\hline

4h & \makecell{$a\ket{000}+b\ket{010}+c\ket{100}$\\$+d\ket{110}$} & No\\
\hline

4i & \makecell{$a\ket{000}+b\ket{010}+c\ket{100}$\\$+d\ket{111}$} & No\\
\hline

4j & \makecell{$a\ket{000}+b\ket{011}+c\ket{101}$\\$+d\ket{110}$} & No\\
\hline

4k & \makecell{$a\ket{000}+b\ket{001}+c\ket{010}$\\$+d\ket{011}$} & \makecell{$\Delta_1'=\Delta_1$\\or $\Delta_1'=\frac{1}{\Delta_1}$}\\
\hline

4l & \makecell{$a\ket{000}+b\ket{001}+c\ket{110}$\\$+d\ket{111}$} & \makecell{$\Delta_1'=\Delta_1$\\or $\Delta_1'=\frac{1}{\Delta_1}$} \\
\hline

4m & \makecell{$a\ket{000}+b\ket{010}+c\ket{101}$\\$+d\ket{111}$} & \makecell{$\Delta_1'=\Delta_1$\\or $\Delta_1'=\frac{1}{\Delta_1}$} \\
\hline

4n & \makecell{$a\ket{000}+b\ket{011}+c\ket{100}$\\$+d\ket{111}$} & \makecell{$\Delta_1'=\Delta_1$\\or $\Delta_1'=\frac{1}{\Delta_1}$} \\
\hline

\end{tabular}
\end{center}
\newpage

$\bullet$ Number of product terms in the state is 5:

\begin{center}
\begin{tabular}{|c|c|c|c|}
\hline
\thead{No.} & \thead{SLICC inequivalet\\states under\\  same no \\of product terms\\ in the state}  & \thead{Existence of \\ infinite SLICC \\inequivalent\\ classes and \\ conditions} \\
\hline

5a & \makecell{$a\ket{000}+b\ket{001}+c\ket{010}$\\$+d\ket{011}+e\ket{100}$}  & \makecell{$\Delta_1'=\Delta_1$}\\
\hline

5b & \makecell{$a\ket{000}+b\ket{001}+c\ket{010}$\\$+d\ket{100}+e\ket{101}$}  & \makecell{$\Delta_5'=\Delta_5$}\\

\hline

5c & \makecell{$a\ket{000}+b\ket{001}+c\ket{010}$\\$+d\ket{100}+e\ket{110}$}  & \makecell{$\Delta_9'=\Delta_9$}\\

\hline

5d & \makecell{$a\ket{000}+b\ket{001}+c\ket{010}$\\$+d\ket{100}+e\ket{111}$}  & \makecell{$\Delta_{11}'=\Delta_{11}$}\\
\hline

5e & \makecell{$a\ket{000}+b\ket{001}+c\ket{010}$\\$+d\ket{101}+e\ket{110}$}  & \makecell{$\Delta_{10}'=\Delta_{10}$}\\
\hline

5f & \makecell{$a\ket{000}+b\ket{001}+c\ket{010}$\\$+d\ket{101}+e\ket{111}$}  & \makecell{$\Delta_9'=\Delta_9$}\\
\hline

5g & \makecell{$a\ket{000}+b\ket{001}+c\ket{010}$\\$+d\ket{110}+e\ket{111}$}  & \makecell{$\Delta_5'=\Delta_5$}\\
\hline

\end{tabular}
\end{center}

$\bullet$ Number of product terms in the state is 6:

\begin{center}
\begin{tabular}{|c|c|c|c|}

\hline
\thead{No.} & \thead{SLICC inequivalent\\states under\\  same no \\of product \\terms in\\the state}  & \thead{Existence of \\ infinite SLICC \\inequivalent\\ classes and \\ conditions} \\
\hline

6a & \makecell{$a\ket{000}+b\ket{001}$\\$+c\ket{010}+d\ket{011}$\\$+e\ket{100}+f\ket{101}$} &  \makecell{$\Delta_1'=\Delta_1 \; and \; \Delta_2'=\Delta_2$\\or\\$ \Delta_1'=1/\Delta_1 \; and \; \Delta_2'=1/\Delta_2$}\\
\hline

6b & \makecell{$a\ket{000}+b\ket{001}$\\$+c\ket{010}+d\ket{011}$\\$+e\ket{100}+f\ket{110}$}  & \makecell{$\Delta_1'=\Delta_1 \; and \; \Delta_3'=\Delta_3$\\or\\$ \Delta_1'=1/\Delta_1 \; and \; \Delta_3'=1/\Delta_3$}\\
\hline

6c & \makecell{$a\ket{000}+b\ket{001}$\\$+c\ket{010}+d\ket{011}$\\$+e\ket{100}+f\ket{111}$}  & \makecell{$\Delta_1'=\Delta_1 \; and \; \Delta_4'=\Delta_4$\\or\\$ \Delta_1'=1/\Delta_1 \; and \; \Delta_4'=1/\Delta_4$}\\
\hline

6d & \makecell{$a\ket{000}+b\ket{001}$\\$+c\ket{010}+d\ket{100}$\\$+e\ket{101}+f\ket{110}$}  & \makecell{$\Delta_5'=\Delta_5 \; and \; \Delta_6'=\Delta_6$\\or\\$ \Delta_5'=1/\Delta_5 \; and \; \Delta_6'=1/\Delta_6$}\\
\hline

6e & \makecell{$a\ket{000}+b\ket{001}$\\$+c\ket{010}+d\ket{100}$\\$+e\ket{101}+f\ket{111}$}  & \makecell{$\Delta_5'=\Delta_5 \; and \; \Delta_3'=\Delta_3$\\or\\$ \Delta_5'=\Delta_5 \; and \; \Delta_3'=1/\Delta_3$}\\
\hline

6f & \makecell{$a\ket{000}+b\ket{001}$\\$+c\ket{010}+d\ket{101}$\\$+e\ket{110}+f\ket{111}$}  & \makecell{$\Delta_2'=\Delta_2 \; and \; \Delta_6'=\Delta_6$}\\
\hline

6g & \makecell{$a\ket{000}+b\ket{001}$\\$+c\ket{011}+d\ket{101}$\\$+e\ket{110}+f\ket{111}$}  & \makecell{$\Delta_2'=\Delta_2 \; and \; \Delta_7'=\Delta_7$\\or\\$ \Delta_2'=1/\Delta_2 \; and \; \Delta_7'=\Delta_7$}\\
\hline

\end{tabular}
\end{center}
\newpage

$\bullet$ Number of product terms in the state is 7:

\begin{center}
\begin{tabular}{|c|c|c|c|}

\hline
\thead{No.} & \thead{SLICC inequivalet\\states under\\  same no \\of product terms\\ in the state}  & \thead{Existence of \\ infinite SLICC \\inequivalent\\ classes and \\ conditions} \\
\hline

7 & \makecell{$a\ket{000}+b\ket{001}+c\ket{010}$\\$+d\ket{011}+e\ket{100}+f\ket{101}$\\$+g\ket{110}$} &  \makecell{$\Delta_1'=\Delta_1 , \Delta_2'=\Delta_2$\\ and $\Delta_8'=\Delta_8$}\\
\hline
\end{tabular}
\end{center}

$\bullet$ Number of product terms in the state is 8:

\begin{center}
\begin{tabular}{|c|c|c|c|}

\hline
\thead{No.} & \thead{SLICC inequivalet\\states under\\  same no \\of product terms\\ in the state}  & \thead{Existence of \\ infinite SLICC \\inequivalent\\ classes and \\ conditions} \\
\hline

8 & \makecell{$a\ket{000}+b\ket{001}+c\ket{010}$\\$+d\ket{011}+e\ket{100}+f\ket{101}$\\$+g\ket{110}+h\ket{111}$} &  \makecell{mentioned\\in appendix}\\
\hline

\end{tabular}
\end{center}

\subsection{Observations from the tables:}

From the tabular form given above, we can summarize classification of pure three qubit states with respect to SLICC in the following way. We will subdivide different SLOCC classes such that each class will contain different SLICC inequivalent classes. In this way, we will try to relate both resource theories for pure three qubit states.\\

\subsubsection{Classification under separable class:}

Firstly, we consider states from the tabular form which lie under separable class, i.e., states of the form $\ket{\psi}^{ABC}= \ket{\phi}^A \otimes \ket{\eta}^B \otimes \ket{\tau}^C $, for which all single qubit reduced systems are pure state. Based on the nature of coherence of reduced systems of the state, we now subdivide this SLOCC class in the following sections.\\

\textbf{All single qubit reduced systems are pure incoherent state:} States within SLICC inequivalent class (1) lie under this section.\\

\textbf{Two single qubit reduced systems are pure incoherent and another one is pure coherent:} SLICC inequivalent classes (2d), (2b), (2a) lie under this. These classes can be divided further into three subsections. If we take states with reduced system A as pure coherent then we obtain class (2d). Similarly, if we take states with reduced system B and C as pure coherent then we obtain classes (2b) and (2a) respectively. So according to the change of coherence nature of particular single qubit systems, we obtain three different SLICC inequivalent classes. \\

\textbf{Two single qubit reduced systems are pure coherent and another one is pure incoherent:} This section contains classes (4k), (4e) and (4h) with condition $ad=bc$. Like before these classes can be subdivided again by taking reduced system A, B and C as pure incoherent respectively. \\

\textbf{ All single qubit reduced reduced systems are pure coherent: }SLICC inequivalent class (8) lies under this section with conditions $\frac{ad}{bc}=\frac{af}{be}=\frac{ah}{bg}=\frac{ag}{ce}=\frac{ah}{cf}=\frac{bh}{df}=\frac{bg}{de}=\frac{ah}{de}=1$.

 \subsubsection{Classification under biseparable class:}

Now we consider states from the tabular form which lie under biseparable class, i.e., states under this class are of the form $\ket{\psi}^{ABC}=\ket{\phi}^A\otimes \ket{\eta}^{BC},\ket{\psi}^{ABC}=\ket{\phi}^B\otimes \ket{\eta}^{AC},\ket{\psi}^{ABC}=\ket{\phi}^C\otimes \ket{\eta}^{AB}$. Like separable class this SLOCC class can be sub classified into following sections and each section contains different SLICC inequivalent classes.\\

\textbf{Two single qubit reduced systems are mixed incoherent and another one is pure incoherent:} Classes (2f), (2e) and (2c) lie under this. By taking reduced system A, B and C as pure incoherent these classes can be classified again into different subsections.\\

\textbf{Two single qubit reduced systems are mixed coherent and another one is pure incoherent:} This section contains classes (3a), (3b), (3f) and (4k), (4e), (4h) with condition $ad \neq bc$. As before by taking any particular single qubit system as pure incoherent these classes lie in different subsections.\\
  
\textbf{ Two single qubit reduced systems are mixed coherent and another one is pure coherent:} Class (8) lies under this. We further notice that if we take reduced system A as pure coherent then we obtain class (8) with condition $\frac{af}{be}=\frac{ag}{ce}=\frac{ah}{de}=1$, similarly, for the choice of reduced systems B and C as pure coherent we obtain class (8) with conditions $\frac{ad}{bc}=\frac{ag}{ce}=\frac{ah}{cf}=1$ and $\frac{ad}{bc}=\frac{af}{be}=\frac{ah}{bg}=1$ respectively. Here we observe that same SLICC inequivalent class lie in different subsections based on different conditions satisfied by the coefficients of the states within it. So, superposition plays crucial role during this subclassification.

\subsubsection{CLassification under genuine tripartite entangled class:}
Now we are considering GHZ and W classes. States within these classes can not be written in separable or biseparable form. These two SLOCC inequivalent classes are genuine entangled classes. Like before we now subclassify these two classes and we want to observe these classes from different point of view.\\

\textbf{All single qubit reduced systems are mixed incoherent states:} Classes (2g), (3e) and (4j) lie under this. We further divided these classes into following subsections. If all bipartite reduced systems are mixture of two incoherent states we get class (2g). If all bipartite reduced systems are mixture of one coherent and one incoherent states we get class (3e) and if all bipartite reduced systems are mixture of two coherent states we obtain class  (4j). So here superposition of bipartite reduced systems play an important role during subclassification. By changing nature of superposition of the reduced bipartite systems we have obtained different SLICC inequivalent classes. \\
 
\textbf{Two single qubit reduced systems are mixed incoherent states and another is mixed coherent:} Classes (3d), (3g), (3c) and (4n), (4m), (4l) lie under this. We now divide these classes in different subsections. If we choose single qubit reduced system A as mixed coherent we get classes (3d) and (4n). Similarly by taking single qubit reduced systems B and C as mixed coherent we get classes (3g), (4m) and (3c), (4l) respectively. Again by changing nature of superposition of reduced bipartite systems classes can be subclassified again. when we consider all bipartite reduced systems are mixture of one coherent and one incoherent states we get classes (3d), (3g) and (3c) whereas by taking bipartite reduced systems as mixture of two coherent states we get classes (4n), (4m) and (4l). Here we observe that nature of superposition of single qubit reduced systems and bipartite reduced systems together play an important role for subclassification. \\

\textbf{ Two single qubit reduced density matrices are mixed coherent states and another one is mixed incoherent:}  By taking reduced systems A, B and C as mixed incoherent we get three different SLICC inequivalent classes (4d), (4g) and (4i) respectively.\\

\textbf{All single qubit reduced systems are mixed coherent:}  We now subdivide classes under this section based on the superposition of the pure states present in reduced systems.

If all single qubit reduced systems are taken as a mixture of one coherent and one incoherent states, the class (4a) appears. 

Now we are considering states from the classes for which two single qubit reduced systems are mixture of one coherent and one incoherent states and another one is mixture of one coherent and two incoherent states. By taking reduced systems A, B and C as mixture of one coherent and two incoherent states the classes (4f), (4c) and (4b) appear respectively. 

If we consider states with two single qubit reduced systems as mixture of two coherent and one incoherent states and another one as mixture of one coherent and one incoherent states we get classes (5a), (5b) and (5c). By taking any particular single qubit system as mixture of one coherent and one incoherent states we can divide again these classes into three different subsections.

By taking all single qubit reduced systems as mixture of one coherent and two incoherent states we get class (5d).

If we consider any one of the single qubit reduced system A, B and C as mixture of two coherent and one incoherent states we obtain classes (5e), (5f), (5g), (6c), (6e) and (6g); where other two reduced single qubit systems of the classes (5e), (5f) and (5g) are mixture of one coherent and two incoherent states and that of the classes (6c), (6e) and (6g) are mixture of two coherent and two incoherent states.

By taking reduced systems A, B and C as mixture of three coherent states we obtain classes (6d), (6b) and (6a) respectively.

If we take all single qubit reduced systems as mixture of two coherent and two incoherent states we get class (6f). 

We now consider class (7)  for which all single qubit reduced systems are mixture of three coherent and one incoherent states.

States within class (8) have the property that all single qubit reduced systems are mixture of four coherent states except the conditions for class (8) mentioned under separable and biseparable classes. 
We have seen that classes under this section can be subdivided depending on the nature of mixture of pure states, i.e., in other words, superposition plays an important role to detect each SLICC inequivalent classes separately.

\textbf{Summarization of overall observations:} We have seen that in overall classification superposition plays an important role. Based on nature of entanglement pure three qubit states can be classified into separable, biseparable, GHZ and W classes. But using superposition of reduced subsystems, it is further possible to subclassify these SLOCC classes. For example, we can take SLICC inequivalent classes (1) and (2a). Both classes lie within same SLOCC class, i.e., separable class but based on nature of superposition of reduced subsystems they lie on different SLICC classes. Same is true for (2g) and (4c) both lie in GHZ class, but based on coherence nature of reduced single qubit subsystems they are two different SLICC classes. So using superposition we are further able to subclassify above SLOCC classes. We have another important observation for the classes (2a), (2b), (2d). Reduced single qubit systems from the states within this class have same property, i.e., two reduced single qubit systems are incoherent and another one is coherent. But based on particular choice of single qubit subsystems they lie in different SLICC classes, i.e., if we interchange nature of coherence of single qubit reduced systems $\ket{\phi}^A$, $\ket{\eta}^B$, $\ket{\tau}^C$ then we get different SLICC classes (2a),(2b),(2d). Same is true for (2f),(2e),(2c) and others. So, in overall classification, superposition of the single qubit reduced subsystems play an important role. Since we have done subclassification under different SLOCC inequivalent classes, so it is possible to detect nature of entanglement through various SLICC inequivalent classes.

\section{Conclusion}

The generalized structure of coherence in multipartite system by characterizing LICC and SLICC inequivalent classes is completely observed. We have provided necessary and sufficient condition for the equivalence of two pure multipartite states under SLICC. We have done complete classification of pure three qubit states using the above condition. We have seen that there exists infinite number of SLICC inequivalent classes for pure three qubit system whereas only six SLOCC inquivalent classes exist for pure three qubit states in resource theory of entanglement. Using superposition nature of reduced subsystems of the states it is possible to  subclassify various SLOCC inequivalent classes, where each subclassified SLOCC class contain SLICC inequivalent classes. It is further possible to detect each SLICC inequivalent classes separately by characterizing reduced single qubit and bipartite systems. Interestingly by interchanging nature of coherence of reduced subsystems we can obtain different SLICC inequivalent classes. So superposition is an important fact for this classification. It is also possible to detect nature of entanglement of pure three qubit states from various SLICC inequivalent classes. In this way we build an better understanding between resource theory of entanglement and resource theory of coherence.
\section*{ACKNOWLEDGEMENTS}
Prabir Kumar Dey acknowledges the support from UGC, India. Dipayan Chakraborty acknowledges the support from CSIR, India. Also, the authors I. Chattopadhyay and D. Sarkar acknowledges the work as part of QUest initiatives by DST India.

\section{Appendix}

$\bullet$ \textbf{Condition for SLICC equivalance of two states obtained by changing coefficients of states} \\

We consider the state where number of product terms 8, if we can convert  $\ket{\psi}=a \ket{000}+ b\ket{001}+c\ket{010}+d\ket{011}+e\ket{100}+f\ket{101}+g\ket{110}+h\ket{111}$ to $\ket{\phi}=a'\ket{000}+b'\ket{001}+c'\ket{010}+d'\ket{011}+e'\ket{100}+f'\ket{101}+g'\ket{110}+h'\ket{111}$ under SLICC then according to above condition there exists SIO operators A,,B,C such that $\ket{\phi}=A \otimes B \otimes C \ket{\psi}$. The required form of A,B,C have been mentioned above.\\

(1) If we convert $\ket{\psi}$ to $\ket{\phi}$ by \[ \; \; \left( \begin{array}{cc}
a_1 & 0 \\
0 & a_2
\end{array}  \right)  \otimes
\left( \begin{array}{cc}
b_1 & 0 \\
0 & b_2
\end{array}  \right)  \otimes
\left( \begin{array}{cc}
c_1 & 0 \\
0 & c_2
\end{array} \right)  
\]
Then we have $\frac{a'd'}{b'c'}=\frac{ad}{bc}$, $\frac{a'f'}{b'e'}=\frac{af}{be}$, $\frac{a'h'}{b'g'}=\frac{ah}{bg}$, $\frac{a'g'}{c'e'}=\frac{ag}{ce}$, $\frac{a'h'}{c'f'}=\frac{ah}{cf}$, $\frac{a'h'}{d'c'}=\frac{ah}{dc}$ \\

(2) If we convert $\ket{\psi}$ to $\ket{\phi}$ by \[ \; \; \left( \begin{array}{cc}
a_1 & 0 \\
0 & a_2
\end{array}  \right)  \otimes
\left( \begin{array}{cc}
b_1 & 0 \\
0 & b_2
\end{array}  \right)  \otimes
\left( \begin{array}{cc}
0 & c_1 \\
c_2 & 0
\end{array} \right)  
\]
Then we have $\frac{b'c'}{a'd'}=\frac{ad}{bc}$, $\frac{b'e'}{a'f'}=\frac{af}{be}$, $\frac{b'g'}{d'e'}=\frac{ah}{cf}$, $\frac{b'g'}{a'h'}=\frac{ah}{bg}$, $\frac{b'h'}{d'f'}=\frac{ag}{ce}$, $\frac{b'g'}{c'f'}=\frac{ah}{ed}$ \\

3) If we convert $\ket{\psi}$ to $\ket{\phi}$ by \[ \; \; \left( \begin{array}{cc}
0 & a_1 \\
a_2 & 0
\end{array}  \right)  \otimes
\left( \begin{array}{cc}
b_1 & 0 \\
0 & b_2
\end{array}  \right)  \otimes
\left( \begin{array}{cc}
c_1 & 0 \\
0 & c_2
\end{array} \right)  
\]
Then we have $\frac{e'h'}{g'f'}=\frac{ad}{bc}$, $\frac{b'e'}{a'f'}=\frac{af}{be}$, $\frac{d'e'}{c'f'}=\frac{ah}{bg}$, $\frac{c'e'}{a'g'}=\frac{ag}{ce}$, $\frac{d'e'}{b'g'}=\frac{ah}{cf}$, $\frac{d'e'}{a'h'}=\frac{ah}{ed}$ \\

4)If we convert $\ket{\psi}$ to $\ket{\phi}$ by \[ \; \; \left( \begin{array}{cc}
a_1 & 0 \\
0 & a_2
\end{array}  \right)  \otimes
\left( \begin{array}{cc}
0 & b_1 \\
b_2 & 0
\end{array}  \right)  \otimes
\left( \begin{array}{cc}
c_1 & 0 \\
0 & c_2
\end{array} \right)  
\]
Then we have $\frac{b'c'}{a'd'}=\frac{ad}{bc}$, $\frac{c'h'}{d'g'}=\frac{af}{be}$, $\frac{c'f'}{d'e'}=\frac{ah}{bg}$, $\frac{c'e'}{a'g'}=\frac{ag}{ce}$, $\frac{c'f'}{a'h'}=\frac{ah}{cf}$, $\frac{c'f'}{b'g'}=\frac{ah}{de}$ \\

5) If we convert $\ket{\psi}$ to $\ket{\phi}$ by \[ \; \; \left( \begin{array}{cc}
0 & a_1 \\
a_2 & 0
\end{array}  \right)  \otimes
\left( \begin{array}{cc}
0 & b_1 \\
b_2 & 0
\end{array}  \right)  \otimes
\left( \begin{array}{cc}
0 & c_1 \\
c_2 & 0
\end{array} \right)  
\]\\
Then we have $\frac{a'h'}{b'g'}=\frac{ah}{bg}$, $\frac{a'h'}{c'f'}=\frac{ah}{cf}$, $\frac{a'h'}{d'e'}=\frac{ah}{de}$, $\frac{a'd'}{b'c'}=\frac{eh}{dg}$, $\frac{a'f'}{b'e'}=\frac{ch}{dg}$, $\frac{a'g'}{c'e'}=\frac{bh}{df}$ \\

6)  If we convert $\ket{\psi}$ to $\ket{\phi}$ by \[ \; \; \left( \begin{array}{cc}
0 & a_1 \\
a_2 & 0
\end{array}  \right)  \otimes
\left( \begin{array}{cc}
0 & b_1 \\
b_2 & 0
\end{array}  \right)  \otimes
\left( \begin{array}{cc}
c_1 & 0 \\
0 & c_2
\end{array} \right)  
\]
Then we have $\frac{a'h'}{b'g'}=\frac{bg}{ah}$, $\frac{b'c'}{a'd'}=\frac{eh}{gf}$, $\frac{a'h'}{c'f'}=\frac{bg}{de}$, $\frac{a'h'}{d'e'}=\frac{bg}{cf}$, $\frac{b'e'}{a'f'}=\frac{ch}{dg}$, $\frac{a'g'}{c'e'}=\frac{ag}{ce}$ \\

7)  If we convert $\ket{\psi}$ to $\ket{\phi}$ by \[ \; \; \left( \begin{array}{cc}
0 & a_1 \\
a_2 & 0
\end{array}  \right)  \otimes
\left( \begin{array}{cc}
0 & b_1 \\
b_2 & 0
\end{array}  \right)  \otimes
\left( \begin{array}{cc}
0 & c_1 \\
c_2 & 0
\end{array} \right)  
\]
Then we have $\frac{a'd'}{b'c'}=\frac{ad}{bc}$, $\frac{c'f'}{a'h'}=\frac{bg}{ae}$, $\frac{c'f'}{b'g'}=\frac{bg}{cf}$, $\frac{b'f'}{a'e'}=\frac{cg}{dh}$, $\frac{c'e'}{a'g'}=\frac{bh}{df}$, $\frac{d'e'}{a'h'}=\frac{ah}{de}$ \\

8) If we convert $\ket{\psi}$ to $\ket{\phi}$ by \[ \; \; \left( \begin{array}{cc}
0 & a_1 \\
a_2 & 0
\end{array}  \right)  \otimes
\left( \begin{array}{cc}
b_1 & 0 \\
0 & b_2
\end{array}  \right)  \otimes
\left( \begin{array}{cc}
0 & c_1 \\
c_2 & 0
\end{array} \right)  
\]
Then we have $\frac{b'c'}{a'd'}=\frac{eh}{fg}$, $\frac{c'e'}{a'g'}=\frac{bh}{af}$, $\frac{a'f'}{b'e'}=\frac{af}{be}$, $\frac{b'g'}{a'h'}=\frac{de}{cf}$, $\frac{c'f'}{a'h'}=\frac{ah}{cf}$, $\frac{d'e'}{a'h'}=\frac{bg}{cf}$ 
\\
We have followed similar methods for states with lower product terms and depending on these conditions we have found existence of infinite number of SLICC inequivalent classes for some states.

\end{document}